\documentclass[sigconf, 10pt, nonacm]{acmart}
\title{Broadcast Rate Limits in Wi-Fi: A Forgotten Bottleneck for Collaborative Edge LLM Inference}

\author{Liujianfu Wang}
\affiliation{%
  \institution{The Chinese University of Hong Kong}
  \city{Hong Kong}
  \country{China}
}

\author{Yuyang Du}
\affiliation{%
  \institution{The Chinese University of Hong Kong}
  \city{Hong Kong}
  \country{China}
}

\author{Shiqi Xu}
\affiliation{%
  \institution{The Chinese University of Hong Kong}
  \city{Hong Kong}
  \country{China}
}

\author{Soung Chang Liew}
\affiliation{%
  \institution{The Chinese University of Hong Kong}
  \city{Hong Kong}
  \country{China}
}
\authornote{Corresponding Author (email: \textit{soung@ie.cuhk.edu.hk}).}

\begin{abstract}
LLM deployment is migrating from data centers to edge devices, where Mixture-of-Experts (MoE) models offer a promising path: sparse expert activation allows the model to be spread across multiple low-cost edge nodes. Distributed MoE inference repeatedly dispatches embeddings from one main node to many workers -- a one-to-many pattern poorly served by the sequential unicasts of mainstream stacks (NCCL, TCP), yet naturally matched by UDP broadcast. We propose a UDP broadcast method for collaborative edge MoE inference, augmented with timeout-driven retransmission 
exploiting near deterministic latency in distributed MoE
for reliability and unordered result gathering for robustness to expert mispredictions, yielding a consistent 1.4× speedup over NCCL and TCP on a wired 8-node cluster. In wireless settings, however, we uncover a deeper, long-forgotten bottleneck: IEEE 802.11 caps broadcast rates at 54 Mbps regardless of physical-layer capacity — a legacy policy built for sparse control traffic, not edge AI.
NS-3 simulations at distances 1m, 2m and 5m show that the optimal rates are much higher (64×, 43×, and 32×, respectively) than the 54 Mbps cap applied in standard.
Thus, we argue that broadcast, is no longer a control-plane relic: it is time for Wi-Fi standards to treat it as a high-throughput data-plane citizen.

\end{abstract}

\begin{document}

\maketitle

\section{Introduction}\label{section_I}
The deployment of Large Language Models (LLMs) is rapidly migrating from centralized data centers to the edge \cite{luo2025toward, li2025tpi}. Driven by strict data privacy requirements and the trend of personal agentic AI, the industry is actively promoting LLM inference on consumer edge devices. For example, Apple is proposing LLM edge deployment on mobile devices by model compression, while Qualcomm is developing toolkits for advanced post-training quantization to fit LLMs on IoT devices \cite{apple2024foundationmodels, qualcomm2023stable, alizadeh2024llm}.

Deploying modern LLMs on edge devices is often hindered by billion-parameter memory footprints that far exceed what edge silicon can accommodate. Mixture-of-Experts (MoE) points to a promising direction for efficient edge LLM deployment by selectively activating a subset of experts in each computation \cite{zhong2024adapmoe, tang2024hobbit}. This sparse activation allows experts to be distributed across multiple nodes, splitting the memory burden across numerous low-cost edge devices.
During model inference, a main node, which hosts non-expert parameters and performs non-expert computations, dispatches embeddings to distributed nodes for parallel expert computations \cite{wang2025od}.
The continuously shuttled embeddings from the main node to multiple expert-hosting nodes result in large one-to-many communication traffic.

\begin{figure}
    \centering
    \includegraphics[width=0.9\linewidth]{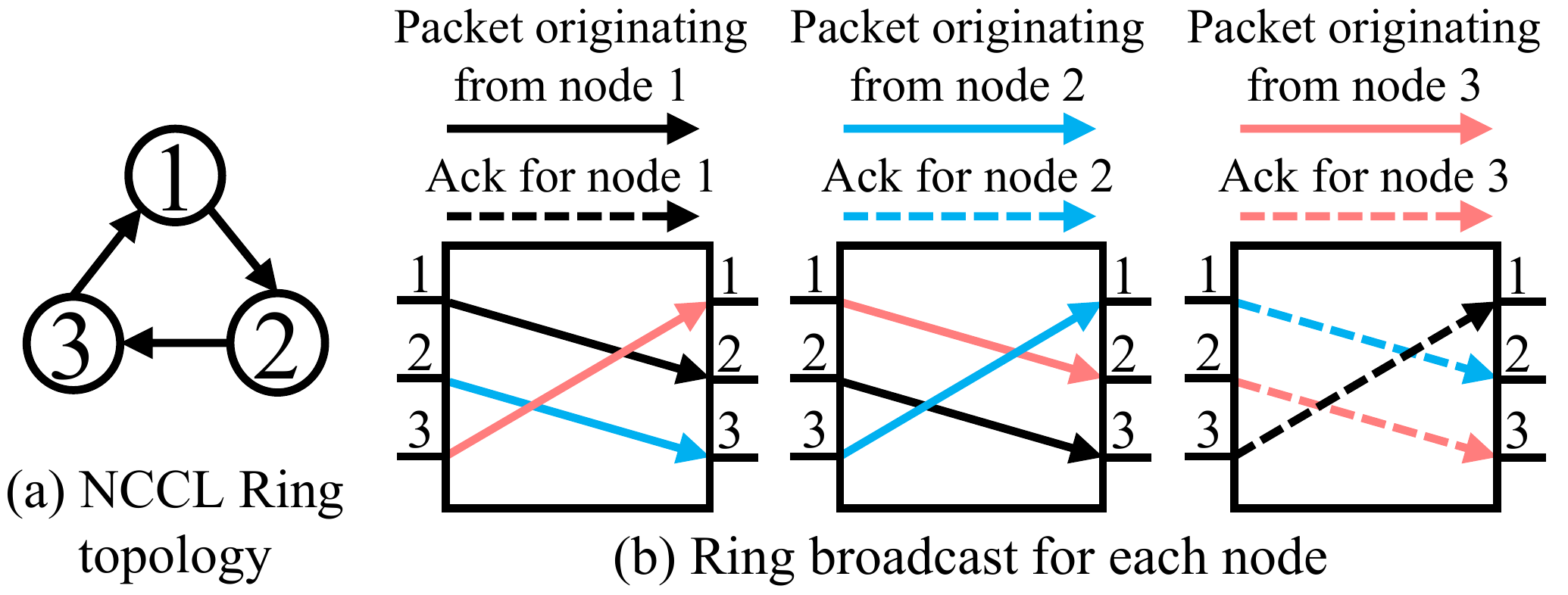}
    \vspace{-0.3cm}
    \caption{(a) NCCL logical ring topology and (b) packet flow in a ring all-to-all broadcast over a full-duplex switch. In each step, each node forwards a message received in the previous step to the next node. A broadcast ends with the last node acking the source node.}
    \vspace{-0.4cm}
    \label{fig:01_nccl_ring_broadcast_only}
\end{figure}

Such an embedding transmission pattern encounters a critical network bottleneck. Popular communication stacks for LLM inference like NCCL \cite{hu2025demystifying} are optimized for \textbf{symmetric all-to-all} broadcasts where all $N$ nodes broadcast simultaneously to all other nodes. NCCL processes these broadcasts as serial unicasts within a logical ring. Fig. \ref{fig:01_nccl_ring_broadcast_only}(a) shows a 3-node setup, in which the logical ring is overlaid over a full-duplex switch interconnecting the nodes.
As illustrated in Fig. \ref{fig:01_nccl_ring_broadcast_only}(b),
altogether $N$ steps are needed  within this ring-based topology.
Owing to the pipelined pattern for the ring topology \cite{hu2025demystifying}, all links connecting to the switch are fully occupied.

However, the traffic pattern in distributed MoE is different: the broadcast traffic is \textbf{asymmetric} and \textbf{one-to-many}: only the main node broadcasts embeddings to expert-hosting nodes. For such a traffic pattern, the NCCL ring broadcast still requires $N$ steps, scaling linearly with node count. Meanwhile, each link is only utilized $1/N$ of the time during the $N$ steps (in fact, only one direction of the full-duplex link), and much bandwidth is wasted. In this scenario, NCCL’s ring-based broadcast offers no advantage over $N-1$ separate unicasts from the main node to other nodes.


The communication pattern where the main node sends embeddings to multiple recipients matches perfectly the one-to-many characteristic of UDP broadcasts, which finishes asymmetric broadcast in one step. For example, an Etherswitch can replicate the packet from the main node and forward copies to all worker nodes in one step. All links connecting to the switch are utilized in that single step. UDP broadcasts remove redundant transfers and slash the load on the source node compared with multiple unicasts. 

However, one important issue inherent in UDP broadcasting is its unreliability. To address this, we augment the classic UDP broadcast with retransmission triggered by a pre-defined timeout.
As the number of activated parameters per expert is fixed, the edge-node computation workload remains stable.
We could set the timeout to be slightly larger than the deterministic computation time, so that a node failing to respond within the timeout is likely to have missed the UDP packet and a retransmission is thus needed. Our revised UDP framework is simple yet highly effective -- in the wired edge cluster tested, it achieves a 1.4× speedup over TCP-backend NCCL, with almost zero additional cost in communication bandwidth.

The wired LAN is only one deployment scenario. As pointed out in \cite{cui2025edge}, many practical edge systems rely on wireless deployments over Wi-Fi for its ubiquity in local environments. Our subsequent experiments reveal an existing limitation in Wi-Fi systems that hinders efficient one-to-many broadcasts. That is, broadcast frames in Wi-Fi are typically confined to a low basic rate ($\leq 54$Mbps), even when the physical capacity of Wi-Fi has grown to 46 Gbps in the latest IEEE 802.11be standard \cite{bellalta2016next, wang2024next}. This restriction made sense historically: since broadcast was meant for the delivery of sporadic discovery and tiny control messages such as SSID beacons, there is no need for high throughput. Rather, reliability in the lack of an ACK mechanism is the issue.

The underlying reason for the design is not physical infeasibility, but the conventional understanding that broadcasts only handle small packets.
The limit is written into the standard to cater for the worst-case scenario in which some nodes may be very far away from the access point. Therefore, a very low data rate is adopted for broadcasts of control packets that need to reach all nodes associated with the access point.
However, for distributed MoE systems deployed over localized indoor environments with limited physical footprints, this is
\textit{overkill} -- for nodes located far from the Wi-Fi router with very weak signal strength, we could exclude such nodes from participating in the MoE system to prevent them from becoming bottlenecks that degrade the overall system performance.

In essence, the optimal broadcast rate involves a trade-off between over-the-air transmission time and packet error rate.
To determine the optimal broadcast rate, we investigate rates under different channel conditions in an NS-3 simulator \cite{baldo2011open}. The results show that, despite diverse channel conditions, the throughput-maximizing bitrate far exceeds the current 54 Mbps ceiling supported by Wi-Fi standards, revealing a MAC design issue for high-throughput broadcast that has long been overlooked.

Driven by the increasingly urgent demand to deploy large foundation models at the network edge \cite{luo2025toward},
distributed MoE inference introduces a new class of workloads where low-latency one-to-many delivery is paramount. Yet, within prevalent consumer edge environments, the restrictive broadcast rates of Wi-Fi severely degrade local inference speeds.
Our simulation results show that achieving optimal efficiency requires much higher rates than the 54 Mbps cap.
We therefore emphasize that broadcast should no longer be viewed as a legacy mechanism solely reserved for sparse control traffic, and urge both industry and academia to reconsider broadcast rate limits for wireless AI infrastructure. By decoupling rate-selection policies from conservative coverage assumptions, we can unlock the true potential of collaborative edge intelligence in practical wireless environments.

\section{Proposed Methodology}\label{section_II}
\subsection{Technical Background}\label{section_II_a}
\textbf{Parameter distribution.} In MoE models, expert parameters dominate the memory footprint, accounting for up to 95\% of GPU memory usage \cite{du2024sida, tang2024hobbit}. However, each individual expert is relatively small. This makes the collaborative edge scenario ideal: while a single edge device has the capacity to host only a limited number of experts or non-expert parameters, the massive deployment of low-cost IoT devices in everyday environments enables the distribution of the full MoE model across multiple devices.

We designate one node as the main node to cache the non-expert parameters, and store expert parameters on distributed nodes that function as workers.
The non-expert computations are done at the main node while expert computations are executed on the worker nodes.
During decoding, expert activations are processed by multiple worker nodes in parallel, with each node hosting a single expert and executing the associated computation task independently.



\textbf{Expert offloading, expert activation prediction, and expert prefetching.} Each worker node has limited GPU memory allowing it to cache only experts likely to be activated. Inactive experts can be offloaded to CPU memory or SSD \cite{tang2024hobbit, zhong2024adapmoe}. This treatment effectively reduces the GPU memory footprint. The catch is that a needed expert may not be cached in the GPU memory and the computation can be stalled while it is loaded from CPU to GPU. An effective solution to minimize such stalls is to predict soon-to-be-activated experts and prefetch them in advance. For example, \cite{eliseev2023fast, tang2024hobbit, zhong2024adapmoe} employs a look-ahead scheme that feeds the latest embedding into gate networks of subsequent k layers to predict expert activations, where k is fixed and pre-defined before execution. With these predictions, worker nodes prefetch the required experts to significantly reduce expert computation stalls due to on-the-fly loading. Both the accuracy and the speed of prediction are critical, as mispredictions or late predictions can still cause significant computation stalls.

\textbf{Embedding transmission.} During inference, communication between the main node and worker nodes occurs in two directions: (1) embedding dispatch from the main node to the worker nodes and (2) the expert-result gathering in the reverse direction. We refer to these processes as downstream and upstream, respectively. For upstream, the workers send different embeddings to the main node, and therefore the upstream consists of multiple distinct unicast data streams, requiring multiple handlings at the main node irrespective of the method used. In contrast, the one-to-many downstream pattern offers opportunities for simultaneous reception and handling of the embedding at multiple worker nodes.

This paper focuses on \textbf{improving downstream efficiency}. The different methods studied in this paper have similar performance in \textbf{upstream efficiency} as they all require $N$ steps to process the results from $N$ workers. 

\subsection{Method}\label{section_II_b}
\textbf{Baseline methods.} 
The methodology section uses the timing diagram of TCP unicast to illustrate the baseline for simplicity. This is because both NCCL and multiple-TCP unicast require $O(N)$ dispatch steps, i.e., TCP effectively captures the core concepts without unnecessary complexity. Experimental benchmarking against NCCL is presented later.

\begin{figure}
    \centering
    \includegraphics[width=0.75\linewidth]{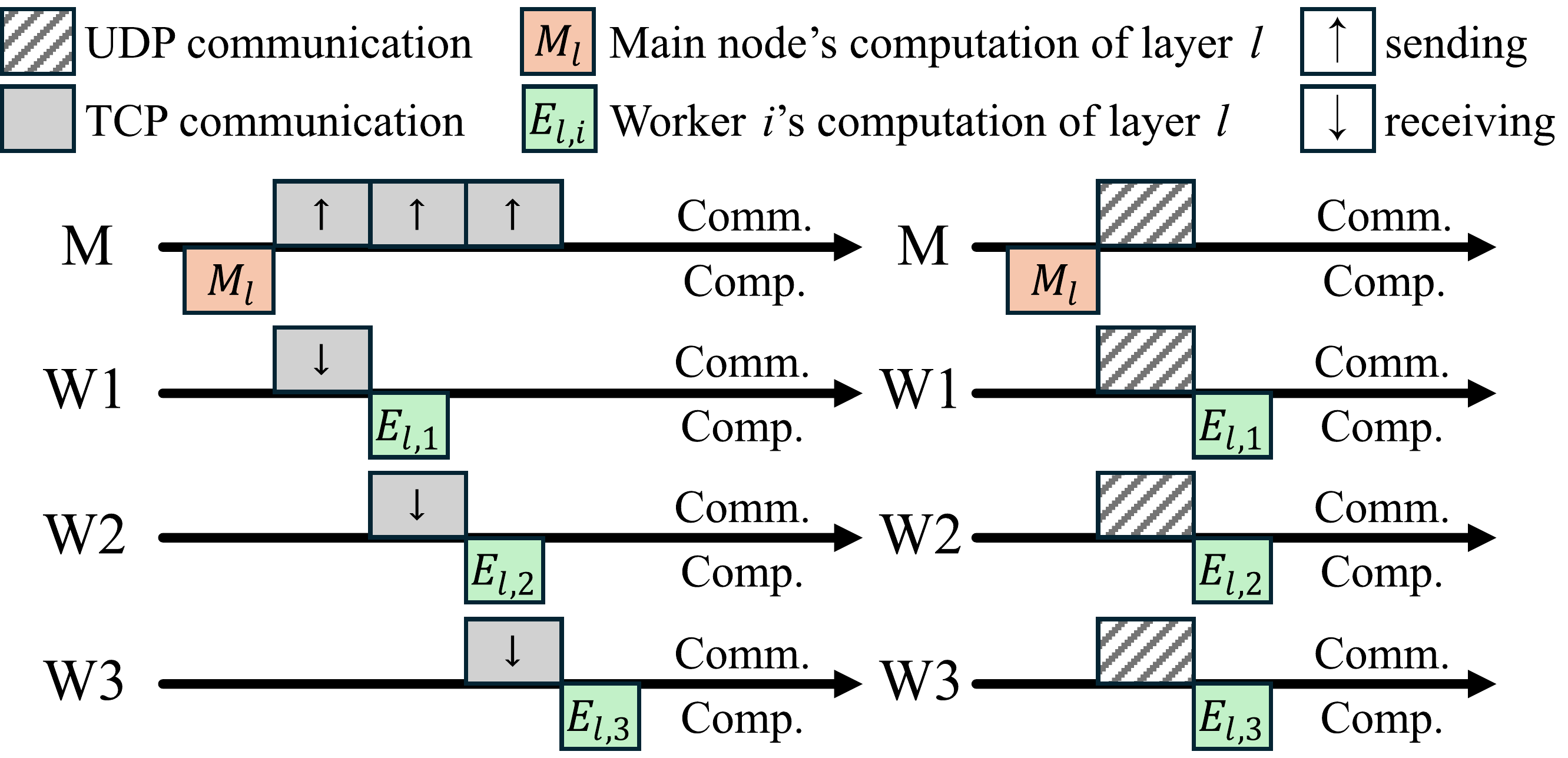}
    \vspace{-0.3cm}
    \caption{Downstream timing diagrams for (a) TCP baseline, (b)our UDP broadcasting method}
    \vspace{-0.4cm}
    \Description{}
    \label{fig:03_tcp_vs_broadcast_no_error}
\end{figure}

In the TCP unicast baseline, as illustrated in Fig. \ref{fig:03_tcp_vs_broadcast_no_error}(a), the main node starts with the non-expert layers and derives the expert activations at the gate layer output. It then dispatches embeddings sequentially\footnote{
Because worker nodes require fully assembled messages to begin computation, parallel TCP communication with multi-threading does not improve overall efficiency as the main node may take turns transmitting packets to different worker nodes (e.g., via round-robin scheduling). In contrast, sequential and contiguous message transmission minimizes delivery latency for early nodes, enabling them to compute and return results sooner. Therefore, this paper adopts sequential rather than parallel TCP.}
to the worker nodes and gathers their results before proceeding to the next layer. For analytical simplicity, we assume each worker processes one expert; more complex mappings do not alter this fundamental transmission pattern. Serial dispatch leads to a longer downstream processing time, and this penalty scales linearly with the number of activated experts. The penalty is particularly concerning as modern MoE models are trending toward activating more experts. For instance, moving from the top-2 expert activation in Mixtral-8x7B \cite{jiang2024mixtral} to the top-8 activation in Qwen3 \cite{yang2025qwen3} and DeepSeek \cite{liu2024deepseek} could quadruple the serial dispatch latency.

\begin{figure}
    \centering
    \includegraphics[width=0.65\linewidth]{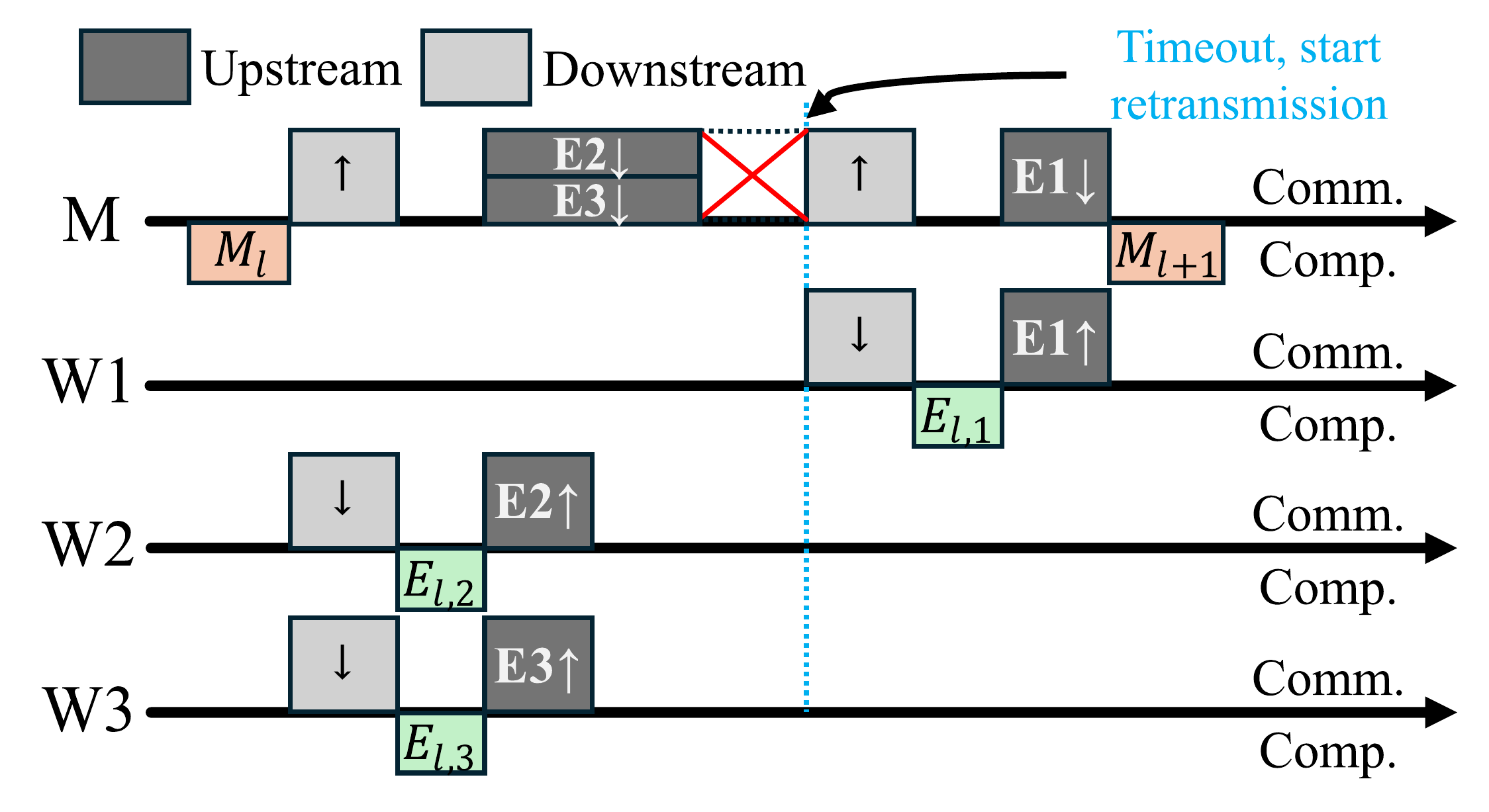}
    \vspace{-0.3cm}
    \caption{Downstream and upstream timing diagram of UDP method with retransmission. 
    \vspace{-0.4cm}
    }
    \Description{}
    \label{fig:04_broadcast_with_retransmission}
\end{figure}

\textbf{Proposed UDP broadcast-based method.} To eliminate the downstream bottleneck, we dispatch embeddings via UDP broadcast. As shown in Fig. \ref{fig:03_tcp_vs_broadcast_no_error}(b), a single broadcast kick-starts all workers in parallel, eliminating the serial-dispatch delay in the downstream transmission. To secure reliability under UDP, we propose an efficient timeout-retransmission scheme that exploits the near deterministic computation latency in the distributed MoE system.
Specifically, since the expert sizes are fixed, and workers’ compute capability remains generally stable, the expected dispatch-and-gather duration can be derived from empirical statistics. By setting a precise timeout threshold based on these statistics, the system can quickly detect lost packets. As shown in Fig. \ref{fig:04_broadcast_with_retransmission}, when a broadcast fails to reach some workers (e.g., Worker 1), the main node falls back to TCP to retransmit the missed packet after the timeout expires.

In the previous examples (i.e., Fig. \ref{fig:03_tcp_vs_broadcast_no_error} and Fig. \ref{fig:04_broadcast_with_retransmission}), we assumed perfect activation prediction, where every required expert was correctly predicted and pre-loaded in the associated worker. In practice, prediction-based prefetching occasionally fails and workers are subject to expert cache misses that incur on-the-fly loading before computation (e.g., Worker 3 in Fig. \ref{fig:05_nccl_with_mispredict} below). These loading events delay the time at which workers can return results to the main node. 

During the gathering process, worker nodes share the link capacity from the EtherSwitch to the main node, making the shared uplink a major bottleneck. If the bottleneck link is not fully utilized and idles while waiting for delayed results (e.g., E3 of Worker 3 in Fig. \ref{fig:05_nccl_with_mispredict}), the overall inference efficiency is compromised. In Fig. \ref{fig:05_nccl_with_mispredict}, the uplink remains idle after completing E1 and E2, until W3 starts to transmit E3 after expert re-loading. This issue stems from the \textbf{ordered} processing mechanism in NCCL, under which basic NCCL primitives follow a fixed order during communication.

\begin{figure}
    \centering
    \includegraphics[width=0.6\linewidth]{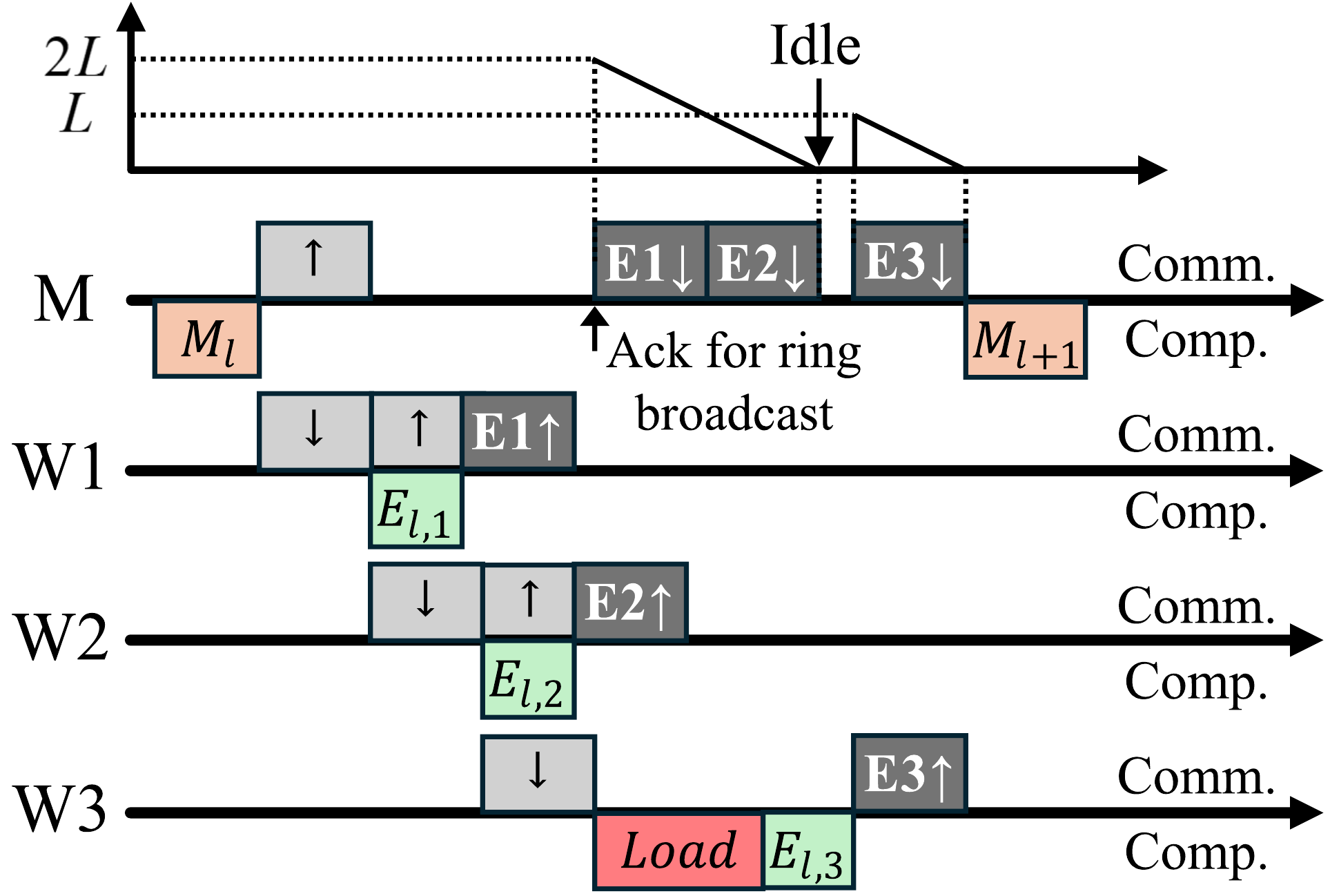}
    \vspace{-0.3cm}
    \caption{Timing diagram of NCCL. The corresponding fluid-flow traffic model (top graph) \cite{liew2010principles_ch8} shows upstream message reception at the main node, modeled as infinitely divisible messages (i.e. fluid). The vertical axis is embeddings (in bits) produced by workers but not yet received by the main node. Message length is $L$.}
    \vspace{-0.4cm}
    \Description{}
    \label{fig:05_nccl_with_mispredict}
\end{figure}

\begin{figure}
    \centering
    \includegraphics[width=1.0\linewidth]{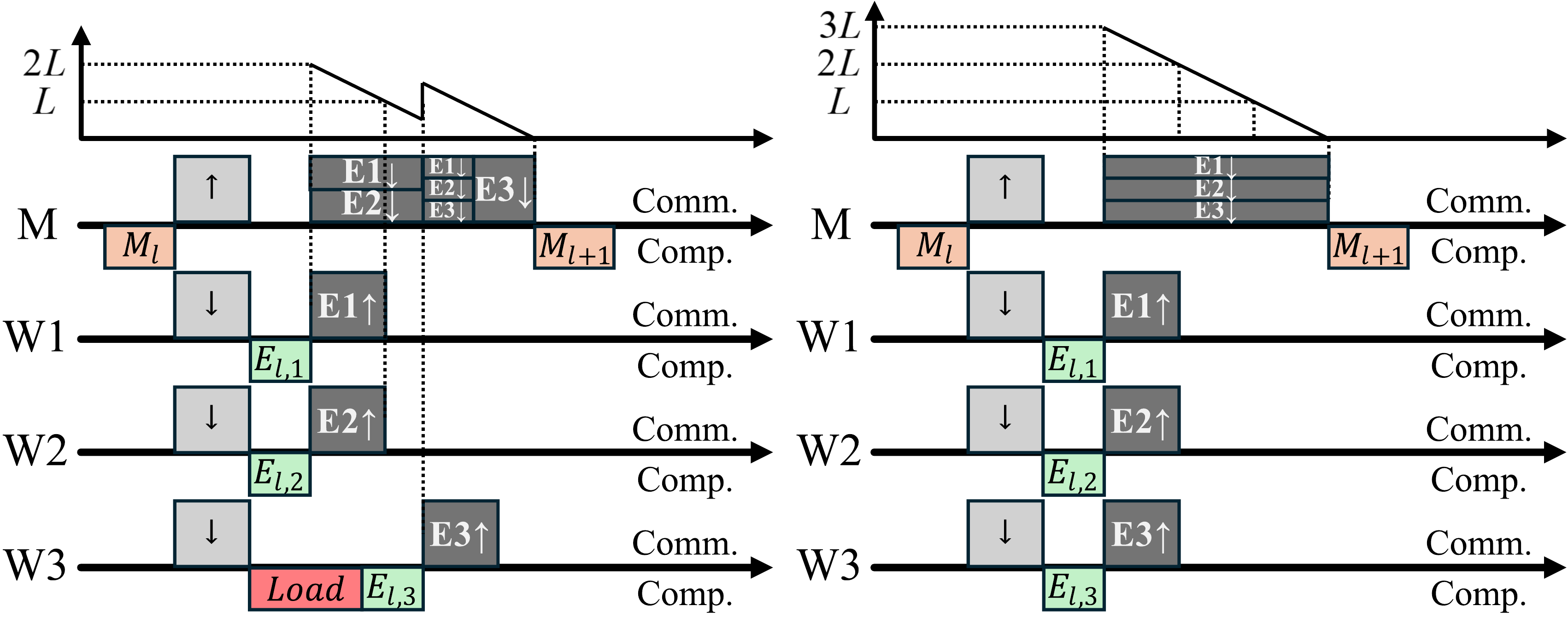}
    \vspace{-0.6cm}
    \caption{Timing diagram of UDP method with corresponding fluid-flow model \cite{liew2010principles_ch8}. (a) with one misprediction v.s. (b) without misprediction.}
    \vspace{-0.4cm}
    \Description{}
    \label{fig:06_broadcast_with_without_mispredict}
\end{figure}

The efficiency degradation problem caused by mispredictions is well addressed in our UDP broadcast method given its \textbf{unordered} gathering nature. That is, the main node collects results from whichever workers return first, granting loading-stalled workers a \textbf{loading margin} to load on-the-fly without sacrificing the overall performance, because the bottleneck uplink is fully occupied by other embeddings during the load and computation time of the loading-stalled workers. As shown in Fig. \ref{fig:06_broadcast_with_without_mispredict}(a), E1 and E2 on the W1 and W2 lines correspond to the “simultaneous” upstream transmissions from workers 1 and 2 on their \textbf{non-overlapping links} from them to the EtherSwitch. However, E1 and E2 need to traverse the \textbf{same link} from the switch to the main node. When W3 finishes the loading and the associated computation, its result begins to be transmitted to the switch. At the switch, this message joins the queue at the bottleneck link from the switch and the main node \textbf{before the queue becomes empty}, and the bottleneck link is fully occupied without idling throughout the uplink transmission process. As a result, the misprediction at W3 does not compromise the efficiency.

Comparing Fig. \ref{fig:06_broadcast_with_without_mispredict}(a) with Fig. \ref{fig:06_broadcast_with_without_mispredict}(b), where perfect prediction is achieved, we note that both scenarios keep the bottleneck uplink busy, thus achieving the same inference efficiency. The two cases demonstrate that unordered gathering enables tolerance to mispredictions without compromising the inference efficiency by keeping the bottleneck link busy.

To quantify the effectiveness of the misprediction tolerance, we define the \textbf{loading margin capacity} as the ability to up to tolerate $X$ mispredictions without performance degradation due to uplink idleness (for example, we have $X=1$ in the above example). The specific value of capacity $X$ depends on the three variables: the expert loading time $T_{load}$, the transmission time of one embedding $T$ ($T=L/R$, where the link bandwidth is $R$ and the message length is $L$, assuming that a unicast and a broadcast take the same amount of time), and the total number of worker nodes $M$. For a UDP broadcast system with $M$ worker nodes, the tolerance capacity $X$ is given by the maximum $x$ such that $T_{load}\leq(M-x)T$.\footnote{Because transmitting $M-x$ correct predictions over the shared link requires $(M-x)T$ time, the system fully utilizes bottleneck capacity and tolerates up to $x$ mispredictions, provided mispredicted workers finish loading before the link idles.} Further experiments about the tolerance capacity are presented in Section \ref{section_III_b}.
\section{Wired Experiments}\label{section_III}

\subsection{Experimental Setup}\label{section_III_a}
\textbf{Base Model and Hardware Setup.} We use Qwen3-30B-A3B \cite{yang2025qwen3} as the base model, which activates 8 out of 128 experts for each token in one decoding layer. The testbed comprises a main node, eight worker nodes, and a prediction node, with the main node and workers each equipped with a single NVIDIA RTX 3090 GPU and are interconnected via a 10 Gbps LAN. This setup represents a typical consumer-grade edge deployment.

\textbf{Expert Prediction Methods Considered.} To evaluate the applicability of our framework across different expert prediction schemes, we consider three different schemes in the prediction node: two configurations of the shadow prediction method from \cite{wang2025od} and the ``NextGate” prediction method from \cite{zhong2024adapmoe}. The shadow prediction method runs a shadow model (quantized model) ahead of the main model and uses its expert activations as predictions. The two configurations tested in the shadow prediction scheme are denoted by Shadow-e4n8 (employing 4-bit experts and 8-bit non-expert layers for the shadow model) and Shadow-e4n16 (4-bit experts and 16-bit non-expert). The NextGate method predicts future expert activations by forwarding the latest embedding vector to the subsequent gating layer. For consistency, all three tested prediction methods run a shadow model on the prediction node. The NextGate method uses the next gate layers of the shadow model (a 4-bit quantized model) for prediction. Note that the communication between the prediction node and the worker nodes only involves transmissions of expert indexes (i.e., a small amount of data), thus the corresponding short communication period is omitted in the timing diagrams presented earlier in this paper.

\textbf{Test Setup and Evaluation Metrics.} We follow the dataset selection from \cite{tang2024hobbit}. To focus on the effectiveness of our design, we quantify the performance of our framework using the per-decoder-layer execution time at the main node, which can be written as $T_{total}=T_A+T_E$. Here, $T_A$ is the total computation time of non-expert layers, including the attention and gate layer computations; $T_E$ is the \textbf{expert layer execution time} as perceived by the main node, which can be further decomposed as $T_E=T_D+T_U+T_C$, where $T_D+T_U$ is the downstream and upstream communication time between the main node and the worker nodes, and $T_C$ is the remaining computation time on the worker nodes, which includes the expert computation time and the potential expert loading time if the mispredictions exceed the tolerance capacity. Correspondingly, the decoding speed can be expressed as $\frac{1}{T_{total}\times N_{layers}}$, where $N_{layers}$ is the number of decoder layers. Our focus will be on the $T_E$ of the different methods, since $T_A$ is the same for all of them.

\subsection{Baseline Comparison}\label{section_III_b}
This subsection benchmarks the $T_E$ of our proposed UDP broadcast-based method against NCCL and TCP baselines across the three prediction methods.

\begin{figure}
    \centering
    \includegraphics[width=1.0\linewidth]{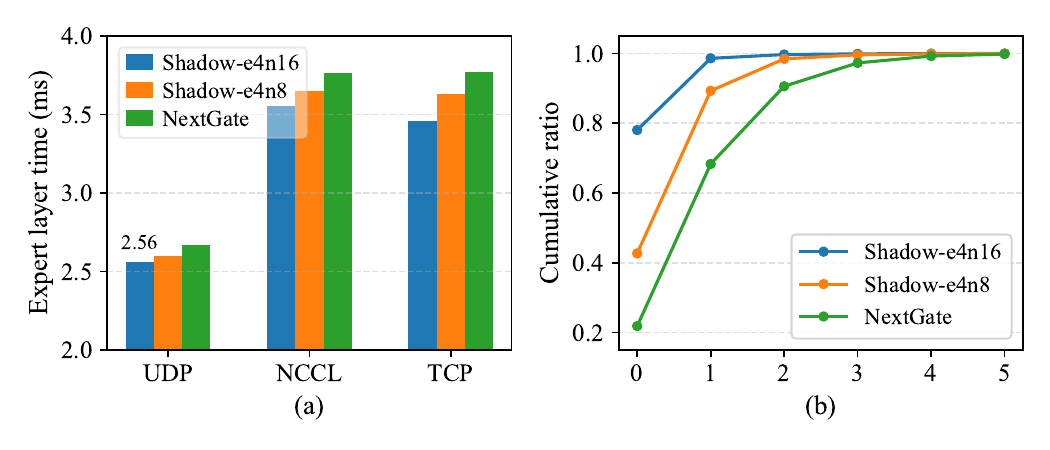}
    \vspace{-0.85cm}
    \caption{(a) Expert layer times across three prediction methods. (b) $\text{Pr}[\#\text{ mispredictions} \leq x]$.}
    \vspace{-0.3cm}
    \label{fig:07_wired_exp}
\end{figure}

Expert layer time. As shown in Fig. \ref{fig:07_wired_exp}(a), our method consistently reduces execution latency compared to NCCL and TCP baselines. For instance, with the Shadow-e4n16 prediction method, our method completed expert layers in 72\% of the time required by the NCCL baseline. This efficiency gain persists with less accurate predictors, achieving an approximate 1.4× speedup over both NCCL and TCP baselines for both Shadow-e4n8 and NextGate methods. The consistent gain over NCCL and TCP baselines across all three prediction methods demonstrates the efficiency of our method.

To contextualize the empirical results, we interpret the latency through a rough analysis. Assuming that the time for both downstream broadcasts and unicasts is $T$, the theoretical speedup of the expert layer time of our method over the baselines can be formulated as: $S = \frac{T_C^{baseline} + T_D^{baseline} + T_U^{baseline}}{T_C^{UDP} + T_D^{UDP} + T_U^{UDP}} = \frac{T_C^{baseline} + 8T + 8T}{T_C^{UDP} + T + 8T}$. Note that $T_C$ varies depending on the specific prediction or communication methods employed. Under the same communication method, $T_C$ increases with the misprediction error rate as mispredictions introduce loading stalls. Across different communication methods, $T_C$ also varies depending on the amount of parallelism. For example, for the TCP baseline method illustrated in Fig. \ref{fig:03_tcp_vs_broadcast_no_error}(a), some of the expert computations overlap with the main node’s transmission and therefore can be discounted from $T_C$, which represents the extra time needed beyond $T_D+T_U$. While this leads to a smaller $T_C$ for the TCP baseline, it comes at the cost of longer downstream transmission times. Conversely, our UDP method as illustrated in Fig. \ref{fig:03_tcp_vs_broadcast_no_error}(b) always incurs the full expert computation time since all workers initiate computation simultaneously at the completion of downstream. 

To estimate the upper bound for the improvement of our method, we look at an idealized system where the computation time is zero (i.e., $T_C=0$) and the system is totally communication-bound. In this case,  the maximum achievable speedup is $16/9 \approx 1.78$. In practice, $T_C^{UDP} \geq T_C^{baseline} > 0$ and the actual speedup observed in our experiments is 1.4. 

\textbf{Robustness under misprediction.} Our design, along with NCCL and TCP methods, shows high tolerance to expert mispredictions. Fig. \ref{fig:07_wired_exp}(b) shows their misprediction CDFs. As shown in Fig. \ref{fig:07_wired_exp}(a), expert layer execution times vary little between the prediction methods despite their differing prediction accuracies, demonstrating the robustness of UDP broadcast, NCCL, and TCP to mispredictions.

\section{Wireless Experiments}\label{section_IV}
\subsection{Experiment Results}\label{section_IV_a}
We next extend our investigation to a wireless environment while maintaining the prediction methods and evaluation metrics in the wired setup. We deploy an ASUS RT-BE96U Wi-Fi 7 router operating at 6 GHz with a 320 MHz bandwidth and four spatial streams. The router connects to the 10 Gbps LAN used in 
Section \ref{section_III}. Unlike the wired experiment, worker nodes now connect wirelessly to the router at \textasciitilde1m via Qualcomm QCNCM865 Wi-Fi 7 NICs with two spatial streams. 

\begin{figure}
    \centering
    \includegraphics[width=0.85\linewidth]{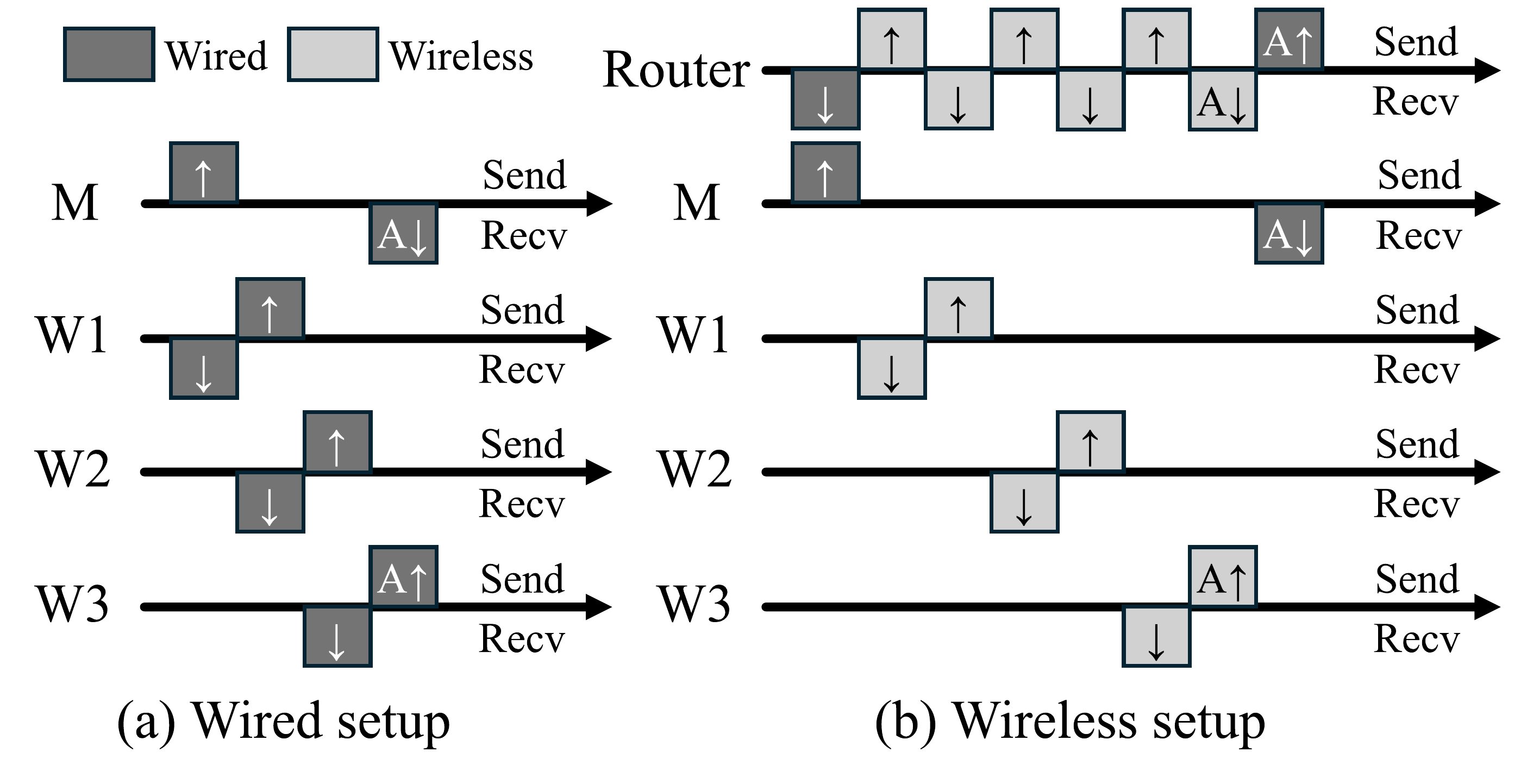}
    \vspace{-0.3cm}
    \caption{Downstream transmission timing diagram of NCCL ring broadcast in (a) wired and (b) wireless setups. "A" denotes acknowledgments.}
    \vspace{-0.3cm}
    \label{fig:08_nccl_issue}
\end{figure}

The shift to the wireless environment significantly alters performance dynamics because the wireless medium is shared and requires all node-to-node traffic to relay through the Access Point (AP). This severely degrades the communication efficiency of NCCL as shown in Fig. \ref{fig:08_nccl_issue}. For NCCL, since each $N-2$ inter-worker unicast needs to be transmitted twice, and the communication from the main node to worker 1 and from worker N-1 to the main node involves two more uses of the wireless medium, the total uses of the wireless medium by NCCL are $2*(N-2)+2 = 2N-2$. In contrast, TCP requires $N-1$ wireless uses and UDP only $1$ use.

\begin{figure}
    \centering
    \includegraphics[width=0.6\linewidth]{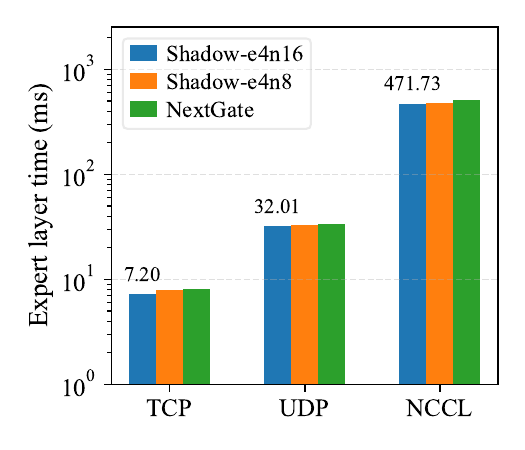}
    \vspace{-0.3cm}
    \caption[The expert layer times in the wireless setup.]{The expert layer times in the wireless setup.}
    \vspace{-0.3cm}
    \label{fig:09_wireless_exp}
\end{figure}

\textbf{Baseline comparison.} Fig. \ref{fig:09_wireless_exp} shows the expert layer times in the wireless setup. The NCCL method's performance degrades dramatically, with expert layer time increasing to 100× that in the wired setup. This degradation stems from both the router relay cost as explained above, and the synchronization overhead of NCCL in an unreliable medium. NCCL primitives (such as broadcast and reduce) are designed assuming stable, high-throughput fabrics and heavily rely on tight data synchronization for efficient communication. However, in unstable wireless environments, channel jitter and packet loss disrupt this tight synchronization, causing severe head-of-line blocking that stalls the entire process. Since wireless scenarios fall outside NCCL’s optimization scope, we focus our wireless analysis on socket-based implementations, namely, the TCP method and our UDP method.

Despite the theoretical advantage analyzed above, our experiments reveal that UDP broadcast incurs a 12.5× penalty compared to its wired performance. The substantial gap between the theoretical efficiency of UDP broadcast and its actual performance can be traced to a legacy protocol design in IEEE 802.11 standards:\textit{ broadcast frames are transmitted at a basic rate capped at 54 Mbps or lower, even if the physical capacity of advanced Wi-Fi standard, such as  Wi-Fi 7, has enabled ultra-efficient communication that supports up to 46 Gbps}. This firmware-enforced limit significantly prolongs the transmission air time, even though the broadcast is only executed once.


Commodity Wi-Fi firmware enforces the basic rate cap and offers no mechanism to override it. To investigate the potential gain of unconstrained high-rate wireless UDP broadcasting, we turn to NS-3 simulations \cite{baldo2011open}.

\textbf{Simulation Workflow.} On the NS-3 simulated links, we simulate the wireless broadcast and unicast transmissions following the dispatch-and-gather communication pattern in our framework, and capture channel statistics such as latency, jitter, and packet error rate at each node. We employ the Groupcast with Retry (GCR) mechanism \cite{daldoul2016performance} for broadcasts, which permits rate selection beyond the basic rate ceiling, and use standard unicasts for the gather phase. We then apply the simulated channel statistics to the wired testbed using traffic control tools, and run the inference system on top of the emulated wireless links to measure the expert layer execution time.

\textbf{Simulation Configuration.} We use NS-3 version 3.45 to simulate the wireless channel conditions as presented in the wireless experiments. Simulated nodes are placed on three concentric circles centered at the router, with radii of 1 m, 2 m, and 5 m, to capture varying link qualities. For the GCR mechanism, we adopt the Unsolicited Retry (GCR-UR) service mode \cite{daldoul2016performance}, and test MCS indices from 0 to 13. The Shadow-e4n16 prediction method is used for the inferences.

\begin{figure}
    \centering
    \includegraphics[width=0.75\linewidth]{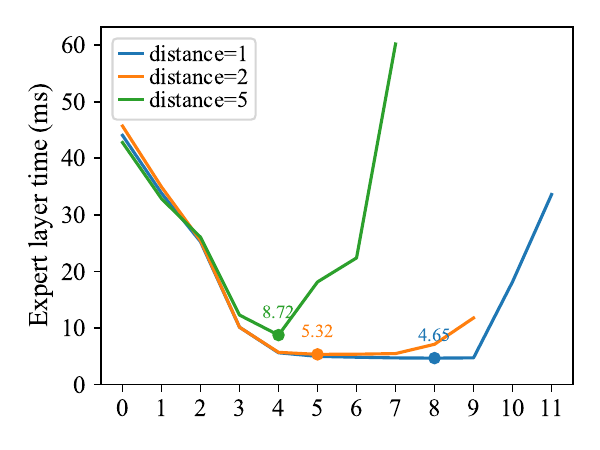}
    \caption{Expert layer time vs. modulation and coding scheme (MCS) index with optimal indexes highlighted.}
    \Description{}
    \label{fig:10_wireless_simulation}
\end{figure}

Fig. \ref{fig:10_wireless_simulation} shows the expert layer execution time across simulated distances and MCS indices. The U-shaped curves in Fig. \ref{fig:10_wireless_simulation} reveal a trade-off: higher MCS indices enable greater bitrates that reduce over-the-air transmission time, but incur higher packet error rates due to denser modulation and coding.
At a distance of 5 m, aggressive modulation schemes cause severe performance degradations due to elevated error rates. Conversely, at closer distances (d = 1 m and 2 m), the execution time plateaus longer before errors outweigh the throughput gains of higher MCS indices.

These results demonstrate that an optimal MCS operating point exists for each distance condition. The optimal MCS indices for distances of 1 m, 2 m, and 5 m are 8 (256-QAM, 3/4 code rate, 3459 Mbps), 5 (64-QAM, 2/3 code rate, 2306 Mbps), and 4 (16-QAM, 3/4 code rate, 1729 Mbps), respectively. The optimal MCS indices increase as the distance decreases, as a shorter distance guarantees a better channel condition and supports a higher optimal MCS value.


At every distance tested, the bitrate at the optimal MCS far exceeds the 54 Mbps ceiling imposed by current Wi-Fi standards. For example, at a distance of 5m, the optimal configuration uses an MCS index of 4 to deliver a bitrate of 1729 Mbps. This rate is 32 times higher than the 54 Mbps limit. This substantial gap highlights the severe bottleneck created by existing firmware limits.

\subsection{Proposed Solution}\label{section_IV_b}
To address the broadcast rate bottleneck while maintaining compatibility with existing network functionalities, we propose decoupling data-plane broadcasts from control-plane broadcasts through the introduction of a \textit{Dual-Mode Broadcast Framework}. Essential network management frames, such as sporadic discovery messages, will continue to operate in legacy mode to ensure compatibility with previous standards and existing devices. For bandwidth-intensive, one-to-many data transfers involving only a subset of nodes with edge LLM inference tasks, the system will employ high-efficiency broadcasting that allows the AP to dynamically select higher MCS indices based on the channel conditions of the targeted nodes.

This hybrid mode design draws on proven principles from existing standards. In IEEE 802.11n, headers use legacy modulation for backward compatibility, allowing older devices to detect channel occupancy and decode critical control information (e.g., NAV). Meanwhile, newer devices leverage high-throughput modulation for the payload to maximize efficiency during the broadcasting process. Our proposal mirrors this approach by preserving compatibility for control-plane messages while optimizing efficiency for LLM inference traffic at the edge.
\section{Conclusion and Outlook}\label{section_V}
We introduce a UDP broadcast framework with retransmission for reliable distributed MoE inference. Specifically, we match UDP broadcast to the MoE one-to-many communication pattern, and augment it with efficient retransmission timeout exploiting the near deterministic communication-computation latency. Moreover, our method allows unordered gathering of expert results, empowering the system with a high expert misprediction tolerance. Our experiments under wired settings show a consistent 1.4× speedup over NCCL and TCP baselines across diverse expert prediction methods.

For wireless deployment, we identify a previously overlooked bottleneck in current Wi-Fi systems: IEEE 802.11 standards cap broadcast frames at a basic rate of 54 Mbps.
Our NS-3 simulations show that 
the optimal bitrates exceed the current ceiling of 54 Mbps by orders of magnitude across distances of 1m, 2m, and 5m.
These findings carry a broader implication. Broadcast rate policies in Wi-Fi were designed for an era of sparse control traffic, not for the bandwidth-intensive one-to-many workloads that distributed edge AI now demands. We advocate that standards bodies and firmware vendors revisit broadcast rate selection, decoupling it from the conservative coverage assumptions for future edge AI development.

\section{Acknowledgements}
\begin{sloppypar}
The work was supported in part by the Hong Kong Innovation and Technology Fund (Project Number: ITS/362/24), and the 2024 Shenzhen-Hong Kong-Macao Science and Technology Program (STIC), Category C (Project Number: SGDX20230821094359004). The experimental work in this paper was conducted in the JC STEM Lab of Advanced Wireless Networks for Mission-Critical Automation and Intelligence funded by The Hong Kong Jockey Club Charities Trust.
\end{sloppypar}

We thank Yuchen Pan for his debugging contribution in building the system. His troubleshooting helps pinpoint bugs in the system.

\bibliographystyle{ACM-Reference-Format}
\bibliography{hotnets26-template}

\end{document}